\documentclass[pdflatex,sn-nature]{sn-jnl}

\usepackage{graphicx}%
\usepackage{amsmath,amssymb,amsfonts}%
\usepackage{booktabs}%
\usepackage[dvipsnames]{xcolor}

\begin{document}

\title[Understanding the mismatch between in-vivo and in-silico rhinomanometry]
{Understanding the mismatch between in-vivo and in-silico rhinomanometry}


\author*[1]{\fnm{Marco} \sur{Atzori}}\email{marco.atzori@polimi.it}
\author[1]{\fnm{Gabriele} \sur{Dini Ciacci}}\email{gab@diniciacci.org}
\author*[1]{\fnm{Maurizio} \sur{Quadrio}}\email{maurizio.quadrio@polimi.it}

\affil*[1]{\orgdiv{Department of Aerospace Science and Technology}, \orgname{Politecnico di Milano}, \orgaddress{\street{Campus Bovisa}, \city{Milano}, \postcode{20156}, \country{Italy}}}

\abstract{
Numerical simulations and clinical measurements of nasal resistance are in quantitative disagreement. Bias introduced by the design of medical devices has not been considered until now as a possible explanation. 
The aim of present paper is to study the effect of the location of the probe on the rhinomanometer that is meant to measure the ambient pressure.
Rhinomanometry is carried out on a 3D silicone model of a patient-specific anatomy; a clinical device and dedicated sensors are employed side-by-side for mutual validation. The same anatomy is also employed for numerical simulations, with approaches spanning a wide range of fidelity levels.
We find that the intrinsic uncertainty of the numerical simulations is of minor importance. To the contrary, the position of the pressure tap intended to acquire the external pressure in the clinical device is crucial, and can cause a mismatch comparable to that generally observed between in-silico and in-vivo rhinomanometry data.
A source of systematic bias may therefore exist in rhinomanometers, designed under the assumption that measurements of the nasal resistance are unaffected by the flow development within the instruments.
}

\keywords{Nasal Resistance,  Rhinomanometry, Nasal Obstruction, Computational Fluid Dynamics}

\maketitle

\section{Introduction}
\label{sec:introduction}
In clinical medicine, nasal airways obstruction (NAO) is a relatively common condition \cite{rhee-etal-2003}, with a polyfactorial etiology \cite{chandra-etal-2009} and a potentially serious impact on a patient's life quality \cite{illum-1997,dinis-haider-2002}, possibly leading to breathing difficulties, sleep apnea, facial pain, dysosmia and other symptoms. Treatments for NAO have a low success rate \cite{dinis-haider-2002}. 
Diagnosis is based upon subjective symptom descriptions and visual inspections of computed-tomography (CT) scans; the main functional information available to the surgeons derives from the clinical exam known as rhinomanometry (RMM), which yields a measure of nasal resistance \cite{demirbas-etal-2011}. 
The nasal resistance quantifies the hydraulic resistance of the nose, defined as the ratio $R=\Delta P/Q$, where $\Delta P$ is the pressure drop across the nasal cavities during breathing, and $Q$ is the flow rate. Higher values of $R$ imply that a stronger exertion is necessary for breathing, suggesting a possible NAO.

\subsection{Challenges in measurements of nasal resistance}
Since precise and non-invasive measurements of the velocity field inside the nose are not possible, rhinomanometry is one application where an in-silico exam could bear substantial advantages.
Several studies have indeed explored the potential of computational fluid dynamics (CFD) for both fundamental research and clinical purposes \cite{quadrio-etal-2014,radulesco-etal-2019}. Simulations could be used, for example, for advanced diagnostics \cite{waldmann-etal-2021}, or to carry out virtual surgeries \cite{radulesco-etal-2020}. 
In order to do so, however, the numerical simulations have to be reliable. 
Numerical studies have reached a progressively higher accuracy \cite[see e.g.][]{calmet-etal-2021}, and can compute values of nasal resistance which are internally consistent \citep{cherobin-etal-2020}. However, a significant mismatch between the CFD computed values and those clinically measured is reported in the literature.
In particular, clinical exams tend to overestimate $R$ w.r.t.\ numerical simulations, even by more than $200-300\%$ \cite{kimbell-etal-2012, osman-etal-2016, berger-etal-2021, schmidt-etal-2022}. 

A conclusive explanation for such large discrepancies has yet to be provided.
Possible reasons investigated so far can be categorized into two classes: i) errors in creating the geometrical representation of the anatomy, and ii) intrinsic simulation errors, caused e.g. by poor turbulence modelling, lack of spatial resolution or inaccurate numerics. 
As an example of the first class, Karbowski \textit{et al.} \cite{karbowski-etal-2023} studied the impact of changing the CT scan segmentation threshold.
For the second class, a precise evaluation of the errors caused by numerical methods, lack of resolution, and turbulent modelling was carried out by Schillaci and Quadrio \cite{schillaci-quadrio-2022}, who concluded that the largest source of error is the order of accuracy of the numerical schemes (an apparently minor detail that is often overlooked). 
Nevertheless, neither poor reconstructions nor numerics can fully explain the vastly different outcomes of CFD and clinical RMM. For a more comprehensive description of this somewhat discomforting state of the art, we refer the reader to the recent review by Johnsen~\cite{johnsen-2024}, who concludes that new possible sources of error should be investigated to reconcile in-vivo and in-silico rhinomanometry. 

\subsection{Aim and scope of the present paper}
This work addresses a source of error that has received little or no attention so far, namely a systematic bias in the in-vivo rhinomanometry caused by the choice of how the pressure difference $\Delta P$ is measured.
The position of the two probes necessary to obtain $\Delta P$ is chosen at the design stage of the clinical device. 
The first probe is placed in a sealed nostril and is supposed to sense the same pressure of the nasopharynx; the other probe is set to measure the ambient pressure. 
Choosing the position of the second probe is apparently not given much attention in the design.

To verify the way $\Delta P$ is evaluated, in this work an experiment is carried out to reproduce in controlled conditions the set up of clinical exams.
The experimental setup reproduces as closely as possible that of an AAR, executed on a silicone phantom model via a clinical device, namely a 4-phase rhinomanometer 4RHINO (Rhinolab, Freiburg, Germany) \citep{vogt-etal-2016}. 
The experiment employs a silicone model of the nasal cavity, which is rigid and not affected by the nasal cycle; the sensors of the clinical rhinomanometer are backed by calibrated laboratory instruments. 
The numerical simulations employ the identical anatomy, thus ruling out any issue with reconstruction accuracy. 
Once the possible sources of error are removed, the uncertainty related to the definition of $\Delta P$ in clinical exams can be meaningfully compared against simulation errors.
Steady inspiration is considered. This is a simplification w.r.t.\ the more complex time-dependent flow rate in human respiration. However, the same simplification is often adopted in simulations to reduce the computational cost \cite{hoerschler-schroeder-meinke-2010}; in the present context, it is useful because it removes part of the possible modelling errors.
The numerical experiments range from the simplest steady-state Reynolds-averaged Navier--Stokes (RANS) simulations, carried out with a commercial code, to state-of-the-art direct numerical simulations (DNS), carried out with a specialized in-house code based on the immersed-boundary method. 
RANS simulations, which by far are the prevailing approach, are performed as in similar studies in the literature, see e.g. \cite{inthavong-etal-2019}.
DNS yields a time-dependent solution where all the active scales are resolved, if the discretization in both space and time is fine enough.
Different meshes for both RANS and DNS are considered. This set of simulations encompasses the broadest range of fidelity levels, and provides a realistic representation of the inaccuracies that can be reasonably expected in numerical simulations.
A smaller second set of simulations is carried out for a test case, representing an idealized junction between a rhinomanometry and the mask worn by the patient. 
The case is designed to highlight how deceptively small differences in instrument design can introduce significant biases.

\section{Methods}

\subsection{Anterior active rhinomanometry}
The experimental setup reproduces as closely as possible that of an anterior active rhinomanometry (AAR) \cite{demirbas-etal-2011}. 
In AAR, the nasal resistances $R_r$ and $R_\ell$ of the left and right airways are measured separately.

During normal breathing, the same pressure difference $\Delta P$ between the external ambient and the nasopharynx exists across both sides of the nasal cavity. During the exam, the patient wears a mask and breaths through the rhinomanometer. Mask and rhinomanometer are connected through a socket that typically also hosts an air filter and the inline flow meter that measures the flow rate. The other ingredient to compute nasal resistance, the pressure drop $\Delta P$, is a pressure difference and thus requires two pressure probes. 

When measuring for example the left resistance $R_\ell$, the right nostril is sealed and a pressure tap is inserted in the sealing. Since there is no flow in the sealed side, the probe through the sealing should sense the pressure in the nasopharynx. The second pressure measurement, that should acquire ambient pressure, is taken with a probe that is typically located in the socket connecting mask and rhinomanometer, just before the filter, to avoid the localized pressure drop across the filter. 
The same process leads to the measurement of $R_r$; an estimate $R_t^\parallel$ of the total nasal resistance $R_t$ is eventually computed as the parallel $R_t^\parallel$ of both, as for the equivalent resistance of two parallel resistors in an electric circuit:
\begin{equation}
  R_t \approx R_t^\parallel = \frac{R_\ell R_r}{R_\ell + R_r} .
\label{eq:parallel}
\end{equation}

\subsection{Anatomy}
The geometry considered in this work for both experimental and numerical measurements is obtained from segmentation at a constant radiodensity threshold of the CT scan of a healthy patient, following standard practices \cite{quadrio-etal-2016, tretiakow-etal-2020}.  
The scan is composed of 384 DICOM images with sagittal and coronal resolution of $0.5 \ mm$ and an axial gap of $0.6 \ mm$. 
The silicone model derived from segmentation, shown in Fig.~\ref{fig:exp}a, is a preliminary version of that employed by Tauwald \textit{et al.} \cite{tauwald-etal-2024} for tomo-PIV measurements of the flow. It is important to mention that, to ease optical access in the PIV experiment, the silicone cast is twice as large as the real anatomy.

\subsection{Laboratory measurements}
A volumetric piston pump is used to create a constant flow rate, in the range between $\approx 100$ and $\approx 600 \, \rm cm^3/s$. The experimental data set includes both inhalation and exhalation, but for brevity only inhalation data are discussed in this paper. 
The accuracy of the internal flow meter of the rhinomanometer is verified via an indirect measurement based on the kinematics of the piston pump, acquired with a high-speed camera. The pressure difference internally measured by the rhinomanometer is backed by independent measurements, employing a GE Druck LPM 9381 transducer (its full scale, accuracy, and sampling frequency are $100 \, \rm Pa$, $0.5 \, \rm Pa$, and $10^3 \, \rm Hz$, respectively). 

\begin{figure}
\centering
\includegraphics[width=0.99\textwidth]{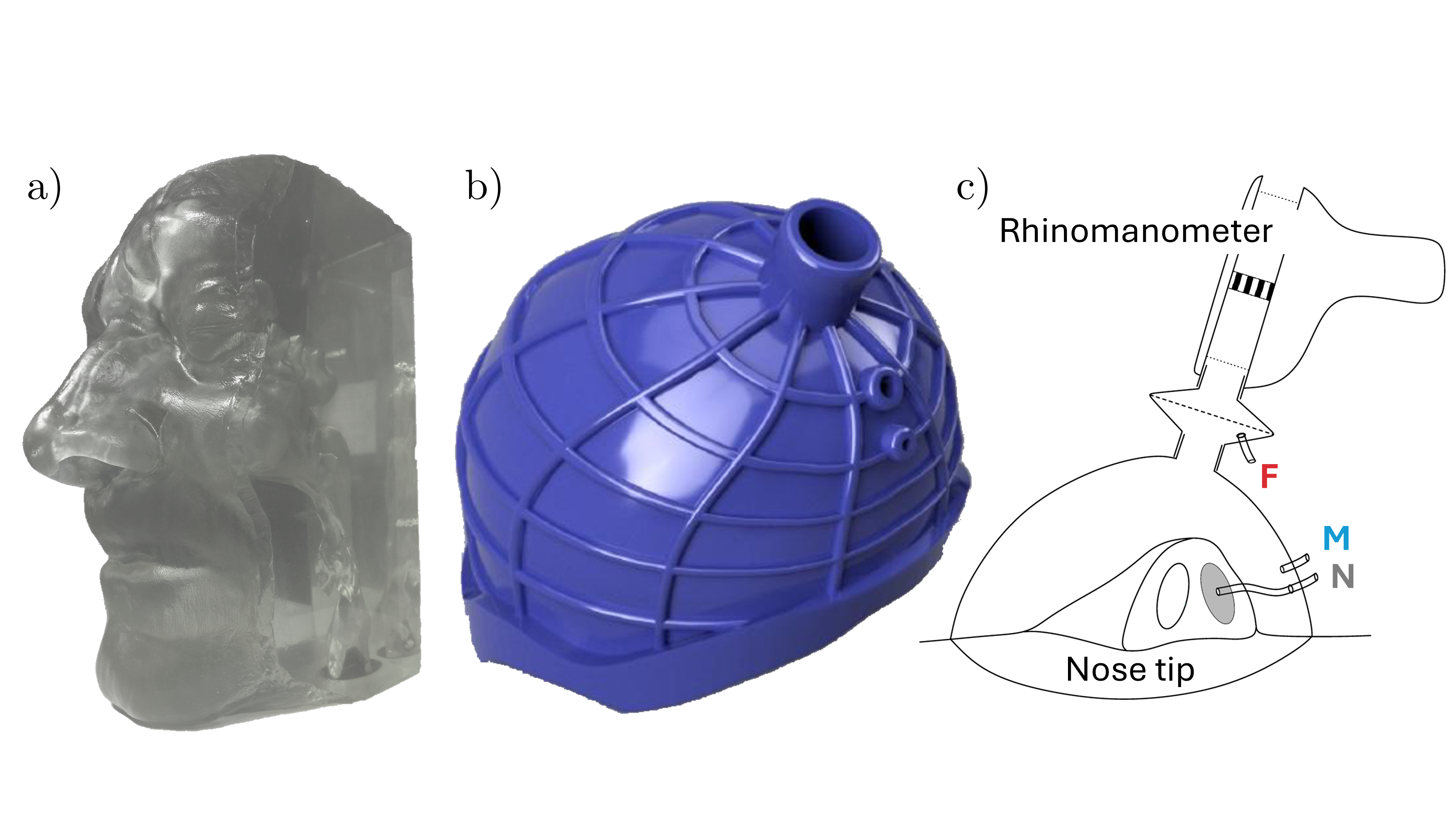}
\caption{a) Silicone phantom and b) 3D-printed mask used in the experiment. c) Sketch of the experimental apparatus. The locations of the three pressure probes are denoted with: sealed nostril ($N$), corresponding to the pressure in the nasopharynx; before the filter ($F$), the typical choice in medical instruments; and at the mask ($M$).}
\label{fig:exp}
\end{figure}

Apart from the 2:1 scale factor of the model, the experiment differs from a clinical AAR exam in terms of the mask. Here, we use the clinical device, but its compliant mask is removed in favour of a custom 3D-printed, semi-rigid mask (shown in Fig.~\ref{fig:exp}b), built to precisely follow the contour of patient's face to avoid leaks, thus removing another important source of uncertainty.

Pressure is measured in three points, sketched in Fig.~\ref{fig:exp}c. One proxy for the ambient pressure, denoted with $F$, is taken in the socket before the filter, as in the clinical device. The second proxy, denoted with $M$, is taken inside the mask and approximately in front of the nostrils. Lastly, the pressure measured with the probe inserted in the sealed nostril and representative of the nasopharynx pressure is denoted with $N$. 

With the experimental setup described above, two measurements of nasal resistance are possible, depending on which pressure probe is chosen to represent the pressure of the outer ambient. There are denoted hereafter with $R_F$ and $R_M$, to indicate that the reference pressure is measured in the filter socket and in the mask, respectively:
\begin{equation}
  R_F = \frac{\Delta P_F}{Q} = \frac{P_N - P_F}{Q} , 
\qquad 
  R_M = \frac{\Delta P_M}{Q} = \frac{P_N - P_M}{Q} .
  \label{eq:RFRM}
\end{equation}

\subsection{Simulations}

\subsubsection{Numerical methods}
The numerical data set produced in this work consists of RANS simulations and high-accuracy direct numerical simulations.
RANS are carried out with the finite-volume code OpenFOAM and its implementation of the $k-\omega$ SST turbulence model \cite{menter-kuntz-langtry-2003}, a popular literature setup \cite{karbowski-etal-2023}. 
Grids are created with OpenFOAM utilities, i.e. \textit{blockMesh} (used to create an initial hexahedral structured grid), and \textit{snappyHexMesh} (that adapts the structured grid to the geometry and refines it). 
The second portion of the data set consists of computationally demanding DNS, carried out with a specialized code for complex geometries based on the efficient immersed-boundary method introduced by Luchini~\cite{luchini-2016, luchini-etal-2025}. 
With the immersed-boundary method, the grid can be a simple Cartesian, uniform grid, that is easy to produce and enjoys uniform quality. These simulations provide an unsteady solution, which is then averaged in time to obtain measurements for the nasal resistance.

\subsubsection{In-silico rhinomanometry}
\begin{figure}
\centering
\includegraphics[width=0.49\textwidth]{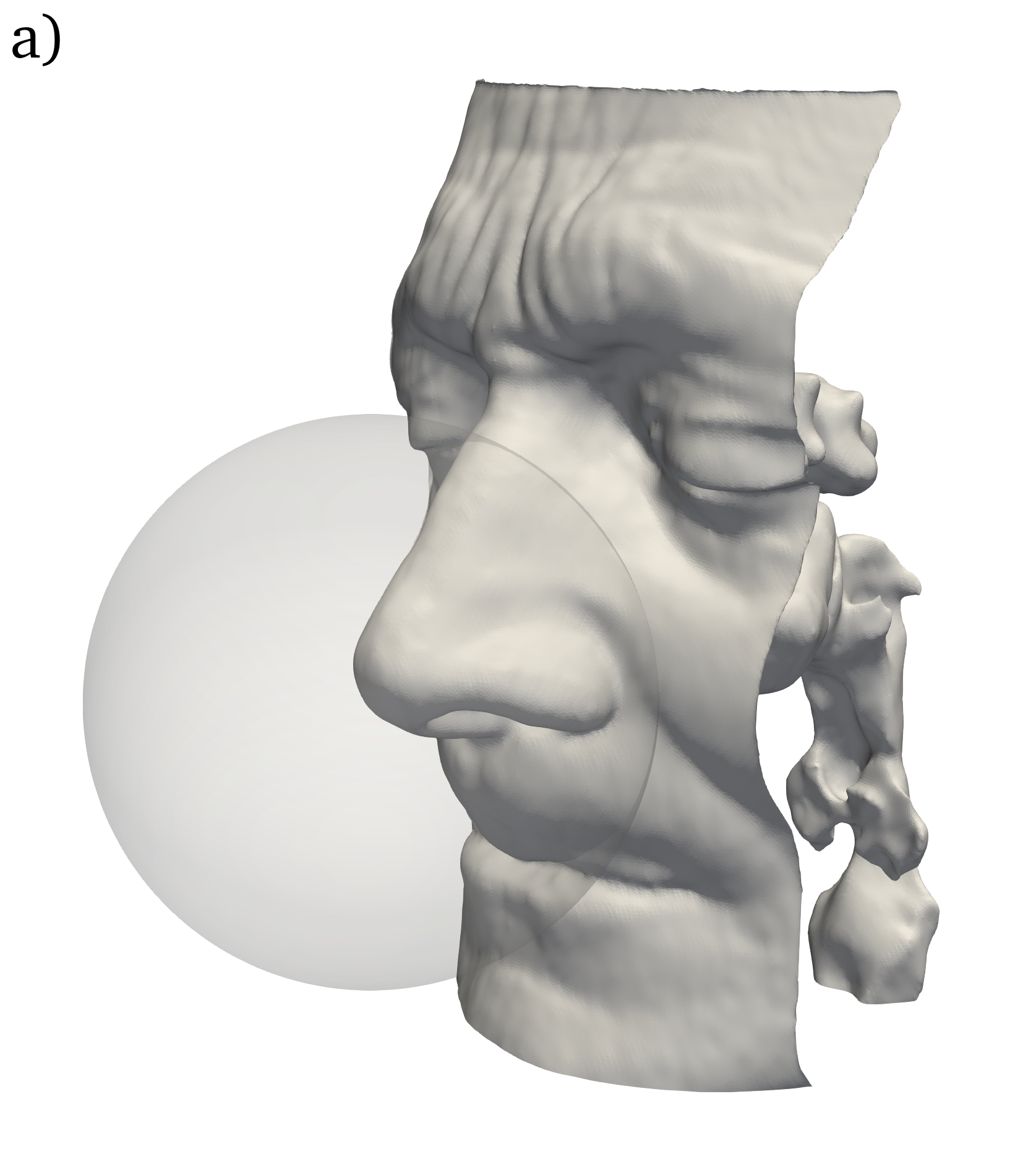}
\includegraphics[width=0.49\textwidth]{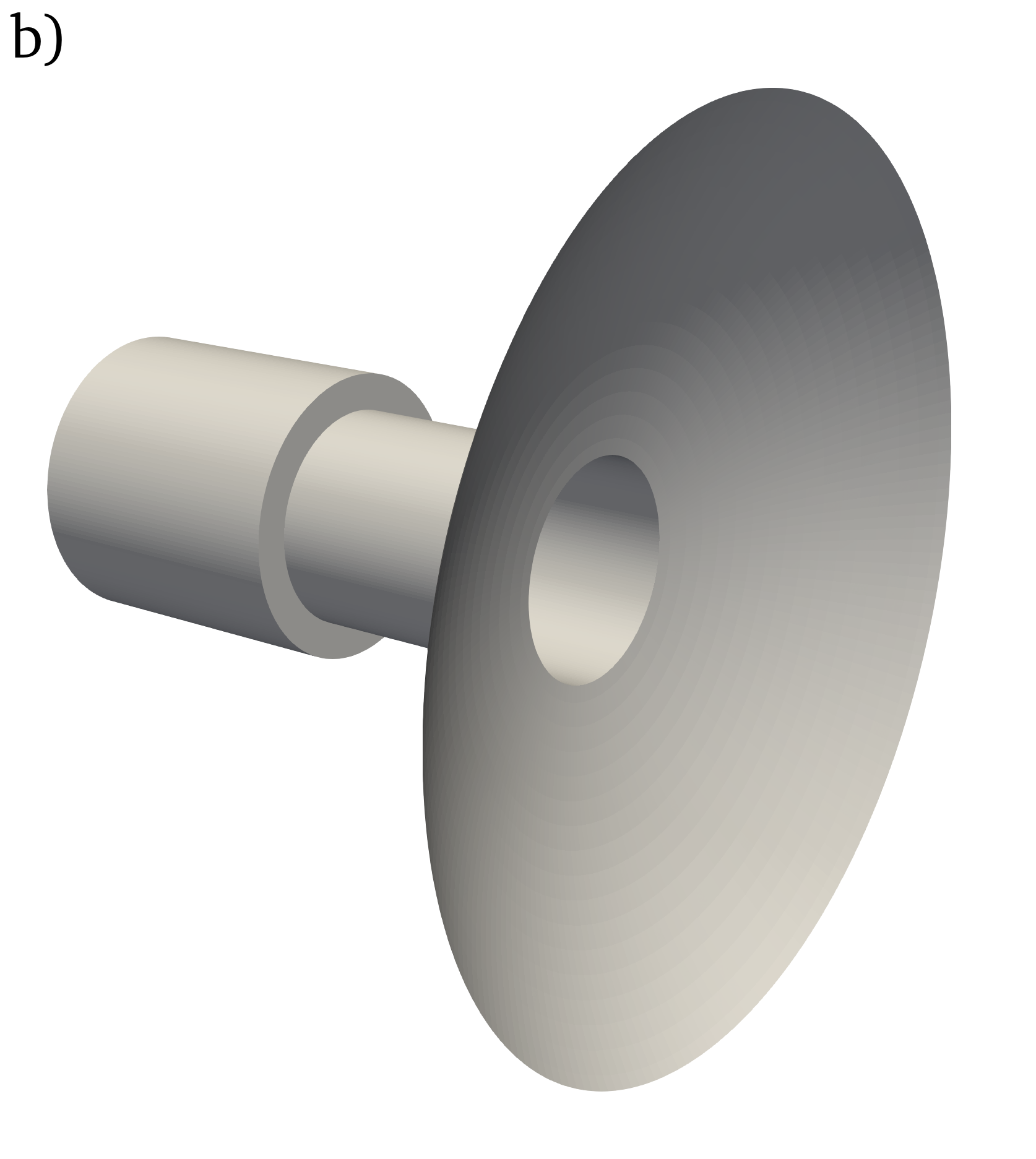}
\caption{Geometry and computational domains of numerical simulations for a) in-silico rhinomanometry, including the outer sphere where inlet boundary conditions are imposed, and b) idealized mask socket.}
\label{fig:simulation_domain}
\end{figure}

The main set of numerical simulations reproduces an in-silico rhinomanometry, using DNS and RANS on the same nasal anatomy used in the experiment. 
The outlet boundary is located below the larynx, whereas the inlet boundary is the surface of a spherical volume around the nose tip (see Fig.~\ref{fig:simulation_domain}a). 
To mimic the experiments, all simulations are carried out for steady boundary conditions. 
In RANS, which directly compute the mean fields, a constant flow rate is imposed. In DNS, a constant $\Delta P$ between inlet and outlet is imposed in each case (note that the values of $\Delta P$ are determined using the RANS simulation, to ensure that both data sets have similar range of flow rates). 
DNS yields an unsteady flow within the nasal cavities or in the pharynx, so that flow rate and pressure drop between nostrils and nasopharynx need to be averaged in time. Time averages are computed by excluding the transient between initial conditions and the statistically stationary regime, discarding up to $0.5 \, \rm s$ of physical time; a further time integration of $1.5 \, \rm s$ is enough to ensure that the error associated with statistical convergence, defined as in \cite{russo-luchini-2017}, remains well below $1\%$. 

\begin{table}
\caption{Total number $N_t$ of grid points and computational cost of simulations at different fidelity levels. CPUh is the number of core-hours required on a conventional CPU core to complete the simulation.}
\label{tab:res_and_cost}
\begin{tabular}{ccc}
\toprule
Simulation type & $N_t$ & CPUh \\
\midrule
RANS (coarse)& $7 \times 10^5$ &   $8$ \\
RANS (middle)& $4 \times 10^6$ &  $44$ \\
RANS (fine)  & $1 \times 10^7$ & $105$ \\
DNS  (coarse)& $3 \times 10^7$ & $100$ \\
DNS  (fine)  & $9 \times 10^7$ & $300$ \\
\botrule
\end{tabular}
\end{table}

Different grids are considered, as summarized in Table~\ref{tab:res_and_cost}, which also reports an estimate of the computational cost of the different methods. Note that the coarsest RANS grid with less than one million cells is at the lower limit of those typically employed for CFD of the nasal cavity, while the third one at ten millions is on the finest side. 
For the DNS, instead, the coarse grid with 30 million points is still slightly under-resolved, whereas the grid with 90 million points is among the largest nasal-flow simulations carried out so far \cite{inthavong-etal-2019}. 
If rescaled to real size, the fine DNS grid has cubic cells with side length of 175 microns, well below the resolution of the CT scan.

For RANS and lower-resolution DNS, two sets of numerical simulations are carried out, by sealing one nostril at a time and measuring the nasal resistance of the other, thus exactly  reproducing the methodology of in-vivo AAR. 
Several flow rates are considered, although the finest DNS grid is used only for an intermediate flow rate. Furthermore, for DNS with the lowest resolution, an additional set of simulations is carried out with both nostrils open, to measure the true nasal resistance directly and compare it with the approximation $R_t^\parallel$ deduced from the monolateral resistances.

Since it has been recently pointed out \cite{schillaci-quadrio-2022} that the single most relevant parameter in a CFD evaluation of the nasal resistance is the order of accuracy of the numerical schemes, we use a second-order discrete form for every differential operator in the RANS equations. However, first-order schemes are used in an additional set of simulations, on the coarsest grid only, to quantify the largest error incurred by the less accurate simulations. The DNS code uses second-order schemes to discretize the spatial derivatives, and a third-order Runge--Kutta scheme for the temporal derivatives.

In terms of computing time, the range extends from the 8 CPU core-hours of the coarsest RANS to the 300 core-hours of the most refined DNS, which includes integration up to $\approx 15 \times 10^3$ time steps, depending on resolution and flow rates. It is noteworthy that the finest RANS and the slightly under-resolved DNS have a comparable computational cost, although the latter includes time integration; this is explained by the overwhelming difference in efficiency between a general-purpose finite-volumes solver and a carefully designed DNS solver.

\subsubsection{Idealized mask junction}
An additional set of simulations focuses on the geometry of an idealized mask junction, to highlight the importance of a seemingly minor detail in the design of a rhinomanometer. The junction is described by assembling three primitive geometrical shapes, namely two cylinders and a spherical cap, as shown in Fig.~\ref{fig:simulation_domain}b). 
The smaller cylinder represents the junction between mask and rhinomanometer. 
The length of both cylinders is fixed at $20 \, \rm mm$, the radius of the larger cylinder is $10 \, \rm mm$, the radius of the spherical cap is $50 \, \rm mm$, and cap is located so that the total length of the geometry is $50 \, \rm mm$. 
The only dimension that is taken as a variable parameter is the radius of smaller cylinder, denoted with $r$. 
The inlet boundary condition, located at the base of the larger cylinder, is an uniform profile, as can be expected downstream the air filter of the rhinomanometer; the outlet boundary condition has a uniform pressure, representing the internal volume of the mask. 
Thanks to the relatively low computational cost to create this data set, only the high-fidelity DNSs are employed here. 

The simplified geometry quantifies the bias introduced in medical devices through the blockage effect caused by a sharp variation in diameter of the internal duct between the air filter and the mask, where the reference pressure probe is often located.
With the notation introduced with Eq.\eqref{eq:RFRM}, the pressure difference measured between the larger cylinder and the outlet directly provides an estimate of the difference between $P_F$ and $P_M$.

\section{Results}

\subsection{General flow features}

\begin{figure}
  \includegraphics[width=0.99\textwidth,trim=75 20 75 50]{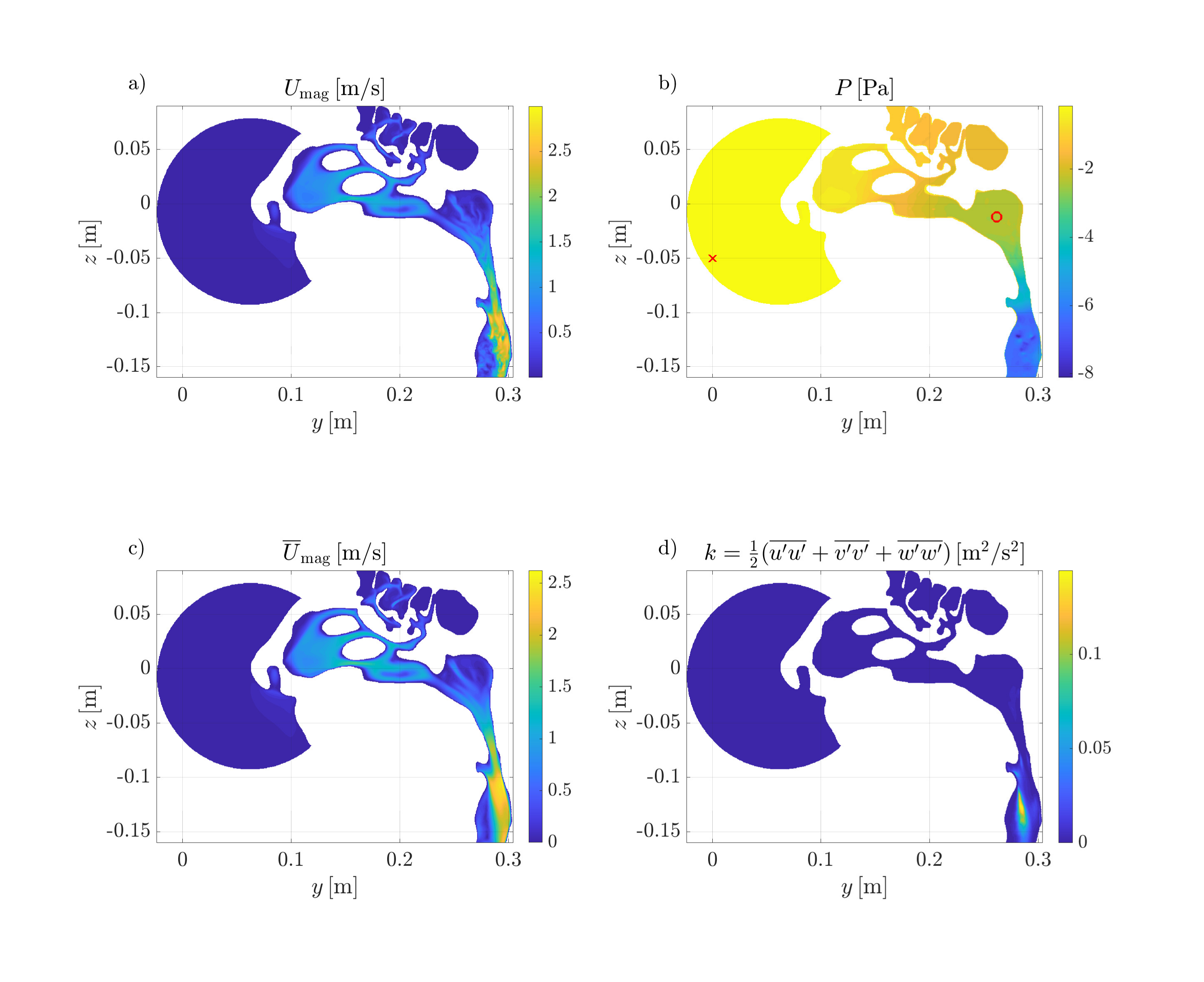}
\caption{Instantaneous (top) and time-averaged (bottom) flow fields on a sagittal section in the DNS simulation with highest flow rate ($Q \approx 630 \rm\,cm^3/s$) and both nostrils open. Panel a): instantaneous velocity magnitude field; b): instantaneous pressure field; c): time-averaged velocity magnitude field; d) time-averaged turbulent kinetic energy field. In panel b), the red symbols indicate where pressure should be ideally extracted to provide a correct representation of the nasal resistance.}
\label{fig:flow}
\end{figure}

Instantaneous and time-averaged flow fields obtained from DNS are shown in Fig.~\ref{fig:flow} in a sagittal section. 
In this case, both nostrils are open and the flow rate is the highest considered, i.e.\ $Q \approx 630\,\rm cm^3/s$. 
Even at this relatively large flow rate, the flow appears to be mostly laminar in the turbinate regions, but becomes unsteady and more irregular in the nasopharynx and in the lower portion of the pharynx, as shown by both the velocity and pressure fields. The turbulent kinetic energy is very small everywhere, being always less than $10^{-3} \, \rm m^2/s^2$ except below the nasopharynx, where it reaches its maximum at almost $0.15 \, \rm m^2/s^2$. In the region in front of the nostrils the air remains mostly at rest, with corresponding negligible pressure differences across the sphere. 
In particular, pressure at the spherical inlet and at the nostril differ by less than $0.1 \, \rm Pa$ only, whereas $\Delta p$  from nostril to nasopharynx is $\approx 2.4 \, \rm Pa$ in this case.
Pressure is also relatively uniform within the nasopharynx, where it varies between $-2.3$ and $-2.5 \, \rm Pa$.
Finally, the laringeal striction induces another significant pressure drop between nasopharynx and outlet, where $P \approx -6 \, \rm Pa$.

\subsection{Discrepancy between simulations at different fidelity levels}
\begin{figure}
\centering
\includegraphics[width=0.75\textwidth]{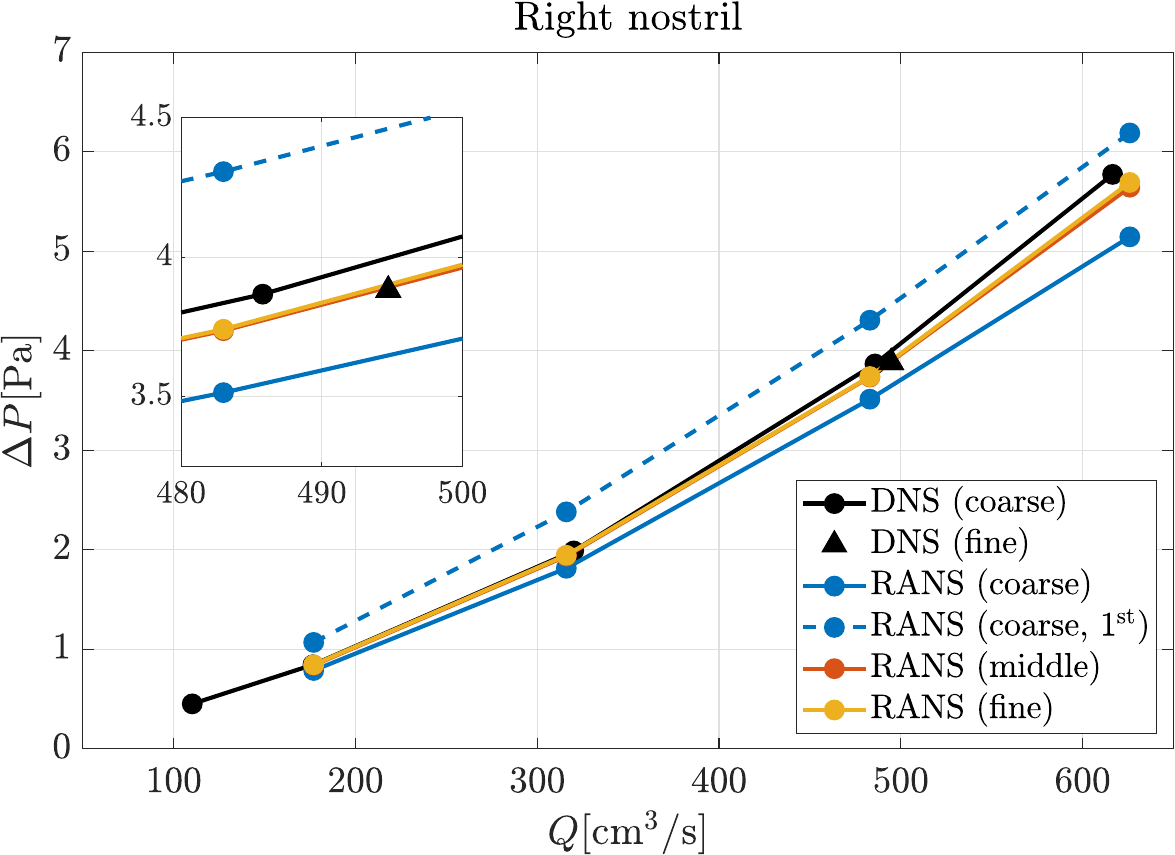}
\caption{Pressure drop between the external ambient and the sealed nostril as a function of the flow rate; results from numerical simulations, left nostril sealed. The indication $1^{\rm st}$ denotes RANS with first-order numerical schemes. Note that the red and orange curves and symbols are mostly overlapping.}
\label{fig:sim}
\end{figure}

We focus for the time being on the pressure difference $\Delta P$ between the external atmosphere and the sealed nostril, as computed from numerical simulations. This is shown in Fig.~\ref{fig:sim} for all the considered fidelity levels as a function of the flow rate; results are plotted for the right nostril (the left nostril yields similar results). 
The coarse RANS is a bit off, but the agreement between intermediate- and higher-resolution RANS is excellent at all flow rates. In particular, the highest discrepancy in the pressure difference, which is found at $Q=624\, \rm cm^3/s$, is of the order of $0.05 \, \rm Pa$, i.e.\ just below $1\%$. For all flow rates lower than $500 \, \rm cm^3/s$, the agreement is even higher, within a $\approx 0.1\%$ error.
At flow rates lower than $Q \approx 300 \, \rm cm^3/s$, these two sets of simulations are also in excellent agreement with the coarser DNS, which yields slightly larger $\Delta P$ at higher $Q$. 
The higher-resolution DNS, with $Q \approx 490\,\rm cm^3/s$ is again in good agreement with the two RANS with higher resolutions. 
Nonetheless, the quantitative discrepancy between the finest and coarsest DNS remains low, at around $2\%$.
Considering the results from all RANS simulations, using coarser grid leads to the opposite effects than in DNS, thus reducing $\Delta P$.
The error introduced with the coarse grid, when a second-order spatial discretization is used, ranges between $7\%$ and $10\%$ from the lowest examined flow rate to the highest. 
Finally, the first-order spatial discretization causes an error in the opposite direction than that produced from under-resolution, and overestimates the pressure difference. 
In the case examined here, this error reaches $17\%$ of $\Delta P$ for the highest flow rate, w.r.t.\ that obtained with the same grid with second-order discretization. 
Overall, the largest discrepancy in $\Delta P$ at a given $Q$ is visible at the highest flow rate, and is of the order of $\pm 10\%$. Once one excludes the two coarsest RANS, which are clearly below the standard, the largest discrepancy shrinks to less than $\pm 3\%$, and is therefore way too small to explain the much larger differences between clinically observed and numerically measured $R$.

\subsection{Discrepancy in $\Delta P$ between simulations and experiments}

\begin{figure}
\centering
\includegraphics[width=0.75\textwidth]{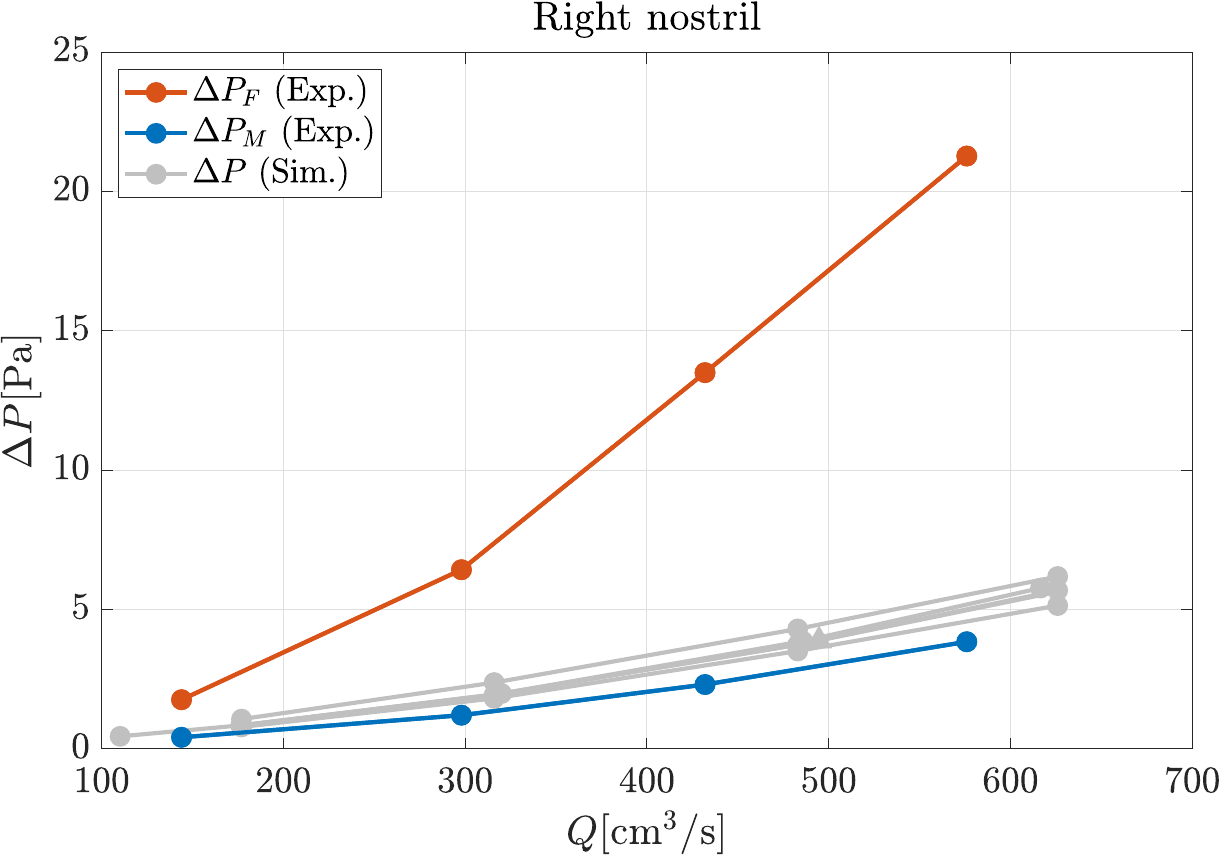}
\caption{Pressure drop between the proxy for ambient pressure and the sealed nostril, obtained from numerical simulations (same data shown in Fig.~\ref{fig:sim}, plotted in gray) and experiments.}
\label{fig:comparison_with_exp}
\end{figure}

Numerically obtained results can now be compared with experimental measurements, as shown in Fig.~\ref{fig:comparison_with_exp}. Both definitions for the experimentally measured pressure drop, namely $\Delta P_M$ and $\Delta P_F$ are included.
Pressure differences derived from numerical simulations are intermediate between $P_M$ and $P_F$, but way closer to the former than to the latter: $\Delta P_F$ is significantly larger than any other estimate, so that the remaining differences are dwarfed.
In particular, $\Delta P_F$ is $\approx 1.3 \, \rm Pa$ higher than $\Delta P_M$ for the lowest flow rate, and $\approx 17 \, \rm Pa$ higher for the highest flow rate. If $\Delta P_M$ is taken as a reference, these values correspond to differences of more than $300\%$ and $400\%$, respectively. 
The quantity $\Delta P_M$ turns out to be much closer to simulation data.
For $Q=576 \, \rm cm^3/s$, for instance, simulations gives an estimate which is around $5 \, \rm Pa$, and $\Delta P_M$ is only $25\%$ lower. 

\begin{table}
\caption{Relative error for the left and right nasal resistances, using the highest fidelity measurement (i.e. DNS on the finest mesh) as a reference.}
\centering
\begin{tabular}{l r r }
&\\ 
\toprule
Side  & \multicolumn{1}{c}{Left} & \multicolumn{1}{c}{Right}  \\
$Q$ & $\approx 414 \rm cm^3/s$  & $ \approx 490 \rm cm^3/s$ \\
\midrule
DNS (coarse) & $ +1.7\%$ & $ +2.6 \%$ \\
\midrule
RANS (fine)    & $-2.2\%$  & $ +0.01 \%$ \\	 	
RANS (middle)    & $-2.4\%$  & $ -0.01 \%$ \\
RANS (coarse)    & $-11.9\%$  & $-6.2 \%$ \\
RANS (coarse, $1^{\rm st}$) & $ +5.5\%$  & $ +14.7 \%$ \\
\midrule
$\Delta P_M$ (Exp)   & $-39.9\%$  & $-24.5 \%$ \\
$\Delta P_F$ (Exp)   & $+168.0\%$  & $ +334.4 \%$ \\
\bottomrule	
\end{tabular}
\label{tab:compareR}
\end{table}

The discrepancy in $\Delta P$ immediately propagates to the nasal resistance. A quantitative estimate of the discrepancy for this quantity, computed for either nostril, is provided in Tab.~\ref{tab:compareR} for the intermediate flow rate, for which the highest-resolution DNS is available and therefore adopted as a reference. 
The nasal resistance for the other cases is interpolated at the corresponding values of $Q$. 
The errors related to the choice of the measurement of the proxy for the outer pressure are found to be above $300\%$. 

Direct comparison between measurements from the rhinomanometer and from the dedicated sensors yields discrepancies of the order of a few percentage points, which is in agreement with the data sheets of the instruments and also orders of magnitude lower than the discrepancies between CFD and clinical exams. For this reason, experimental errors are not considered further on in this paper, and the readings from the internal sensors are just assumed correct.

\subsection{Discrepancy in $R_t$ between simulations and experiments}

\begin{figure}
\centering
\includegraphics[width=0.75\textwidth]{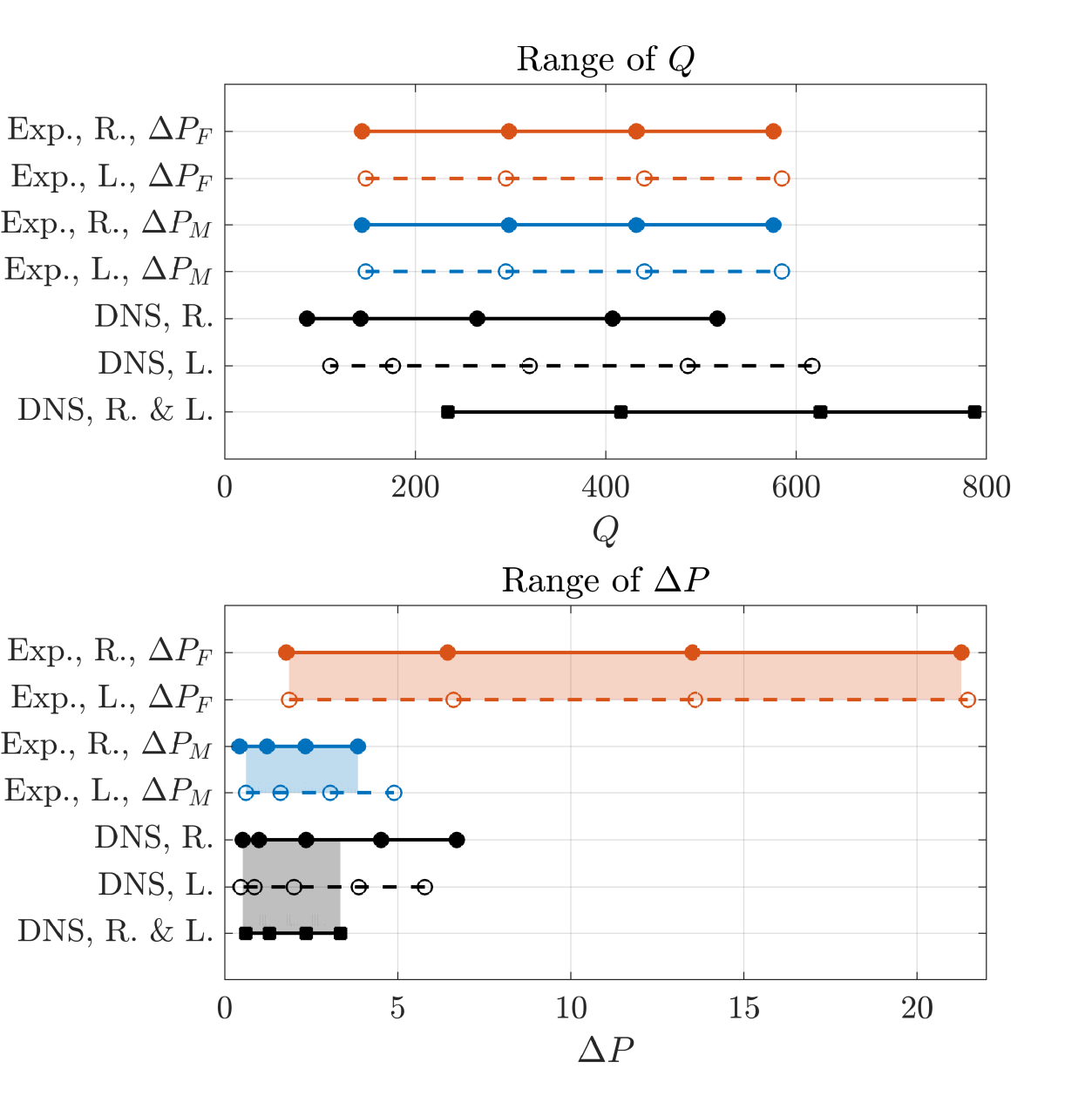}
\caption{Ranges of (top) flow rates and (bottom) pressure differences in each series of data. The open nostrils are denoted with R. and L. The shaded rectangles indicate the range for which $R^\parallel_t$ and $R_t$ are compared hereafter.}
\label{fig:comparison_of_R_range}
\end{figure}

\begin{table}
\caption{Ranges of flow rate and pressure differences in the each series of data.}
\label{tab:comparison_of_R_range}
\begin{tabular}{c c c}
	&\\ 
	\toprule
	Data set & Range of $Q\,[\rm cm^3/s]$ & Range of $\Delta P\,[\rm Pa]$ \\
	\midrule
	Exp, right nostril ($\Delta P=P_{SN}-P_{F}$) & $[144,576]$ & $[1.77,21.27]$ \\
	Exp, left nostril ($\Delta P=P_{SN}-P_{F}$) & $[147,585]$ & $[1.85,21.46]$ \\
	\midrule
	Exp, right nostril ($\Delta P=P_{SN}-P_{M}$) & $[144,576]$ & $[0.42,3.85]$\\
	Exp, left nostril ($\Delta P=P_{SN}-P_{M}$) & $[147,585]$ &$[0.60,4.89]$ \\
	\midrule
	DNS (3e7), right nostril & $[86,517]$ & $[0.51,6.70]$  \\	 	
	DNS (3e7), left nostril & $[110,617]$ & $[0.45,5.77]$  \\	 	
	\midrule
	DNS (3e7), both nostrils & $[234,787]$ & $[0.60,3.33]$  \\	 	
	\bottomrule
\end{tabular}
\end{table}

Values of total resistances are now compared for the experimental and lower-resolution DNS data set, including the simulations with both nostrils open. 
Note that measuring the true $R_t$ directly is only possible with simulations. For the experiments, as in clinical AAR, the estimate $R_t^\parallel$ defined as in eq.~\eqref{eq:parallel} is used. 
From an operative point of view, since eq.~\eqref{eq:parallel} holds for fixed $\Delta P$, computing $R_t^\parallel$ requires interpolating the curves $R_r(\Delta P)$ and $R_\ell(\Delta P)$ to a common set of $\Delta P$ values. 
A more fundamental consequence of this definition of nasal resistance, however, is that discrepancies in $\Delta P$ such as the ones reported above also imply a variation of the intervals where $R_t^\parallel$ can be defined. Table~\ref{tab:comparison_of_R_range} and
Fig.~\ref{fig:comparison_of_R_range} illustrate the ranges of $Q$ and $\Delta P$ for each data set, as well as the intervals where $R_t$ and $R_t^\parallel$ are evaluated.

\begin{figure}
\centering
\includegraphics[width=0.75\textwidth]{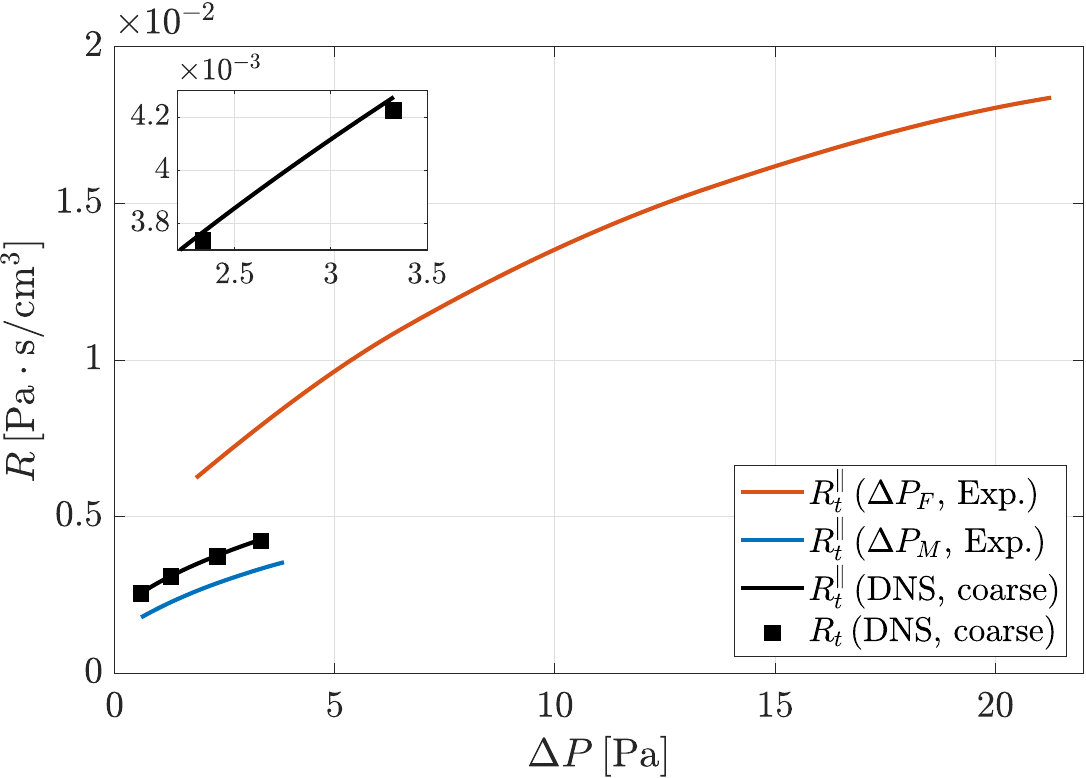}
\caption{Comparison between true total nasal resistance $R_t$ from simulations with both nostrils open and total resistance $R_t^\parallel$ obtained (for simulations and experiments) with the parallel formula \eqref{eq:parallel}, as in clinical exams.}
\label{fig:comparison_of_R}
\end{figure}


The total nasal resistance, computed using the interpolated lateral resistances $R_\ell$ and $R_r$ when necessary, is shown in Fig.~\ref{fig:comparison_of_R}. A quantitative measure of the relative error implied by the parallel approximation of Eq. \eqref{eq:parallel} is provided by using the highest-resolution dataset computed with both nostrils open. The values of $R_t^\parallel$ and $R_t$ differ by $1.2\%$ only, and quantify for the first time the minor inaccuracy involved in the parallel-resistance formula.

Obviously, as expected from the observations on $\Delta P$, the experimental measurements of $R_t^\parallel$ based on $P_M$ are in much better agreement with numerical results than those based on $P_F$. 
By using the same DNS-derived reference, and by considering the highest value of $\Delta P$ for which $R_t^\parallel$ is defined for numerical simulations ($\Delta P=3.22 \, \rm Pa$), together with the range of nasal resistance for each data set, the experimental estimate $\Delta P_M$ is off by $-20.9\%$, whereas $\Delta P_F$ is off by $+87.4\%$.

\subsection{Resistance of the mask junction}
\begin{figure}
\centering
\includegraphics[width=0.99\textwidth]{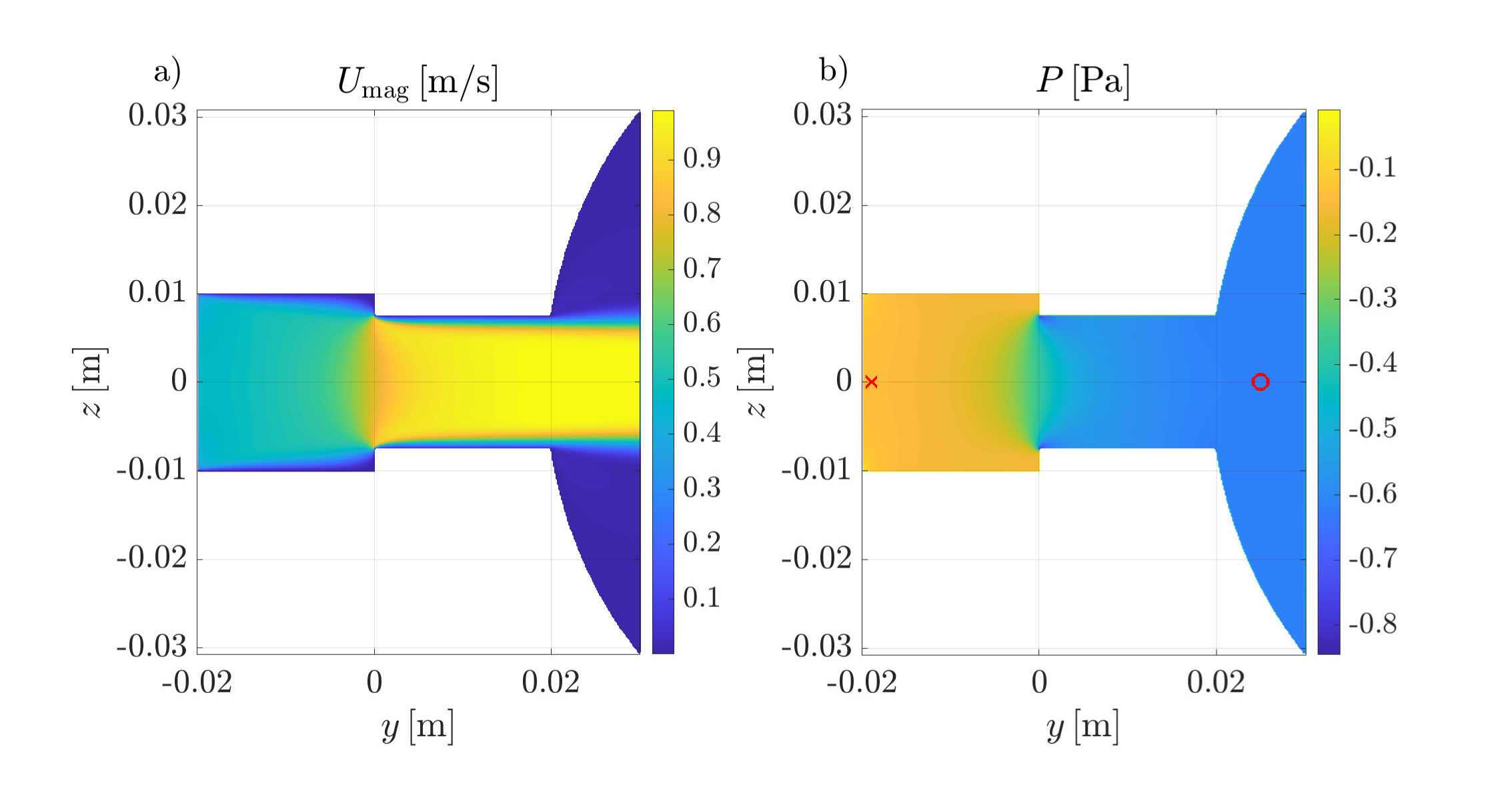}
\caption{Instantaneous velocity magnitude (panel a) and pressure (panel b) fields on a sagittal section for idealized mask junction with $r=7.5 \,\rm mm$. The red cross and circle denote the locations of pressure measurement.}
\label{fig:socket}
\end{figure}

\begin{figure}
\centering
\includegraphics[width=0.75\textwidth]{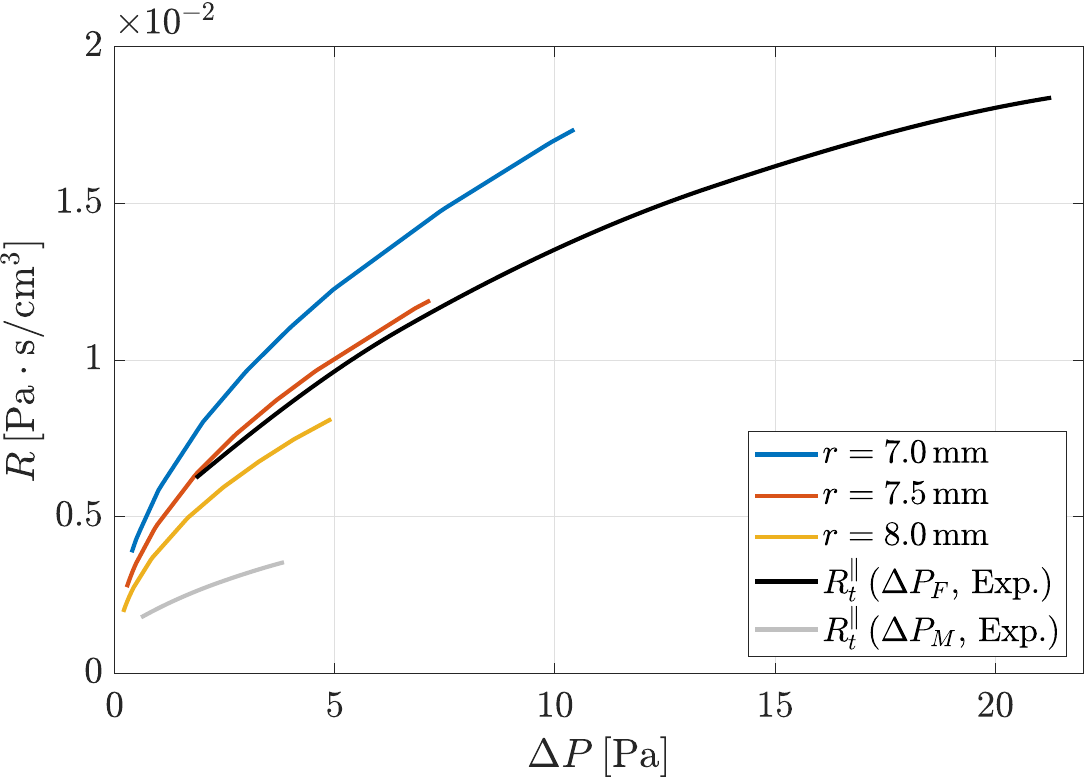}
\caption{Comparison between resistance of the idealized mask junction for three values of parameter $r$ and the experimental measurements of total nasal resistance (same data shown in Fig.~\ref{fig:comparison_of_R}). }
\label{fig:comparison_mask_junction}
\end{figure}

We now examine the resistance of the idealized mask junction, where the geometrical parameters are representative of an actual mask. An overview of the flow is shown in Fig.~\ref{fig:socket}, for a case where the smaller/larger cylinders have radii of $r=7.5 \, \rm mm$ $r=10 \, \rm mm$, and for $Q\approx 100 \, \rm cm^3/s$, when the flow is steady.
The pressure drop between the position of the inlet and the outlet of this test case, i.e. between the outlet of the filter and the inside volume of the mask, is almost entirely caused by the sudden contraction at the mask junction.
The hydraulic resistance for this test case for three values of the smaller radius, i.e.\ $r=7.0 \, \rm mm$, $7.5 \, \rm mm$, and $8.0 \, \rm mm$ is compared against the total nasal resistance in Fig.~\ref{fig:comparison_mask_junction} for the range of $\Delta P$ corresponding to flow rates between $100$ and $600 \, \rm cm^3/s$ through the mask junction. 
For all cases, including when the differences between external and internal radius is minimal at $2 \, \rm mm$ only, the resistance of the idealized mask junction is larger than that measured in the experiment using $\Delta P_M$ as reference pressure.

\section{Discussion}
Our preliminary observations from flow visualizations obtained with high-fidelity numerical simulations corroborate the general principles of how the nasal resistance is measured.
Measuring negligible pressure differences between the inlet of the computational domain and the nostril confirms that, in such numerical simulations, the ambient reference pressure can be safely measured at any point located reasonably far from the nostrils. 
Similarly, pressure within the nasopharynx region is rather uniform, despite the relatively complex flow pattern there, and corresponds well to the static pressure measured at the sealed nostril during an AAR procedure, therefore confirmed to be correct.

We found that simulation uncertainties are led, as indicated in Ref. \cite{schillaci-quadrio-2022}, by the order of accuracy of the numerical schemes, whose importance in this specific application overwhelms, perhaps counter-intuitively, other effects like turbulence modelling. 
DNS presents higher sensitivity to a change of resolution w.r.t.\ RANS, which is the expected consequence of the lack of turbulence modelling in the former. 
Although a dedicated grid-sensitivity analysis would be necessary to assess a possible residual uncertainty on the finest-grid DNS at high flow rates, the associated uncertainty remains small. 
Since the numerical data produced in this work include the two extrema of well-resolved, high-fidelity DNS and coarse-mesh, low-fidelity RANS, an upper bound to the typical errors in measuring nasal resistance computationally for the case study is provided. 
This bound, together with the design of the present study which eliminated other sources of uncertainty (e.g. reconstruction), rules out the possibility that the disagreement between measured and computed values of nasal resistance can be attributed to limitations of the computational model.

The spatial location where the ambient pressure is measured appears to be important for in-vivo rhinomanometry; its choice produce effects that overwhelm errors in numerical simulations. 
In particular, the significant differences between nasal resistances obtained with the two reference pressures measured in the experiments (i.e.\ in the mask, or just before the filter) is a strong indication that the pressure drop within the measuring device itself should not be neglected. 
These effects are large enough to explain the mismatch between CFD and clinical exams.
A faithful reproduction of the entire experimental set up with high-fidelity simulations has not been attempted yet, but the study of an idealized mask junction performed here shows that one possible source for bias is a contraction in the duct between the position of reference pressure measurements and the mask.
In a realistic scenario, a reduction of cross section can occur because of either a sudden change of the internal radius, or an obstruction. For instance, in certain rhinomanometer designs, the duct additionally contains the flexible silicone tube that connects the device to the sealed nostril. 
Viscous losses caused by skin friction alone, owing to the relatively short distance that must exist between the position of pressure measurement and the nostril, are not sufficient to account for the discrepancy. 

Only one anatomy and (more importantly) one combination of rhinomanometer and mask has been considered in our study. Moreover, it should also be remembered that the quantitative outcome of this study could be influenced by the size of the silicone model, enlarged by a factor of 2. However, the relative agreement among clinically measured resistances is suggestive of similarities in the design of rhinomanometers, at least in terms of the position of pressure probes.

The reasons why the importance of the pressure probe position has not been noticed so far are left to speculation. Existing CFD studies measure the nasal resistance directly, rather than reproducing the whole set up of in-vivo rhinomanometry with a mask. In fact, in-vitro rhinomanometry, which is in better agreement with simulations than in-vivo rhinomanometry (see e.g. Cherobin \textit{et al.}~\cite{cherobin-etal-2020}), also employs 3D replicas that are limited to the nasal cavity, and do not employ a mask.
Moreover, the complexity of the nasal cavity, with its narrow spaces, may induce the false impression that the corresponding resistance is relatively large, leading to overlooking the importance of properly choosing the reference pressure.

\section{Conclusions}

Motivated by the substantial disagreement between the clinically measured and the numerically computed nasal resistance $R$, in this work a systematic campaign of combined numerical and laboratory experiments has been carried out, designing methods and procedures to minimize the effects of several sources of errors in the measurement of $R$. 
In agreement with the existing literature, comparing simulations with different fidelity levels has confirmed that intrinsic uncertainties related to numerical simulations do not come close to explaining the mismatch between in-silico and in-vivo rhinomanometry. 
This conclusion still holds true, even when simulations run with quite coarse resolutions are considered.

We have been able to verify some of the hypotheses at the root of AAR, as for example the estimate of the true nasal resistance via the parallel resistance formula \eqref{eq:parallel}.
However, the main outcome of this study is the indication that a source of systematic error may exist in the design of the rhinomanometers typically used in clinical practice.
In particular, the measured nasal resistance is highly sensitive to the positioning of the pressure tap that senses the ambient pressure. 
In numerical simulations, there is no need for a mask, and ambient pressure is naturally available and detectable at some distance from the nostrils; similarly, in-vitro experiments can easily measure ambient pressure. In-vivo rhinomanometry, instead, involves a mask worn by the patient. The mask exit is connected to the rhinomanometer proper via a socket, and this creates what effectively works as an artificial extension of the nasal cavity, with an associated resistance that is far from negligible. 
In other words, the ambient pressure, when measured at some location along the way from the nostrils to the rhinomanometer filter, is bound to depend, and significantly so, on the geometrical details of the mask and its junction with the rhinomanometer, and on the probe position as well.

Further studies are definitely needed to properly explain quantitatively the discrepancy between measured and computed $R$, and to arrive at the design of improved devices for clinical use. However, we are confident that the present contribution highlights an important physical aspect that has not been given proper attention so far.

\subsection*{Acknowledgements}
The Authors are grateful for the support by the ENT group of ss.Paolo e Carlo University Hospital, Università degli Studi di Milano, who provided and assisted in using the rhinomanometer employed in the study.
This research has been partially supported by ICSC--Centro Nazionale di Ricerca in High Performance Computing, Big Data, and Quantum Computing funded by European Union: NextGenerationEU. MQ acknowledges support from the PRIN 2022 project OpenNOSE, cod.2022BYA5AF CUP D53D2300343006. 
Computing time was provided by the CINECA Italian Supercomputing Center through the ISCRA-B projects ONOSE-AN and ONOSE-AI. 

\section*{Declarations}

\subsection*{Conflict of interest} The authors declare that they have no potential
financial conflicts of interest related to the research.

\subsection*{Short Bio of the Authors} 

{\bf Marco Atzori} obtained a master’s degree in physics at the University of Genoa and a doctorate in engineering mechanics at the Royal Institute of Technology (KTH). He is now a researcher at Politecnico di Milano and studies the flow in the human lungs.

{\bf Gabriele Dini Ciacci} has a master’s degree in Aerospace Engineering at Politecnico di Milano where he was a research assistant. He is a chartered engineer, inventor and admin of an engineering company.

{\bf Maurizio Quadrio} is full professor of Fluid Mechanics at Politecnico di Milano. His research interests include turbulence, and the fluid mechanics of the human body. He leads the {\em OpenNOSE} project.


\begin{thebibliography}{10}
\expandafter\ifx\csname url\endcsname\relax
  \def\url#1{\burl{#1}}\fi
\expandafter\ifx\csname urlprefix\endcsname\relax\def\urlprefix{URL }\fi
\providecommand{\bibinfo}[2]{#2}
\providecommand{\eprint}[2][]{\url{#2}}
\providecommand{\doi}[1]{\url{https://doi.org/#1}}
\bibcommenthead

\bibitem{rhee-etal-2003}
\bibinfo{author}{Rhee, J.}, \bibinfo{author}{Book, D.},
  \bibinfo{author}{Burzynski, M.} \& \bibinfo{author}{Smith, T.}
\newblock \bibinfo{title}{Quality of {{Life Assessment}} in {{Nasal Airway
  Obstruction}}}.
\newblock \emph{\bibinfo{journal}{The Laryngoscope}}
  \textbf{\bibinfo{volume}{113}}, \bibinfo{pages}{1118--1122}
  (\bibinfo{year}{2003}).

\bibitem{chandra-etal-2009}
\bibinfo{author}{Chandra, R.}, \bibinfo{author}{Patadia, M.} \&
  \bibinfo{author}{Raviv, J.}
\newblock \bibinfo{title}{Diagnosis of {{Nasal Airway Obstruction}}}.
\newblock \emph{\bibinfo{journal}{Otolaryngologic Clinics of North America}}
  \textbf{\bibinfo{volume}{42}}, \bibinfo{pages}{207--225}
  (\bibinfo{year}{2009}).

\bibitem{illum-1997}
\bibinfo{author}{Illum, P.}
\newblock \bibinfo{title}{Septoplasty and compensatory inferior turbinate
  hypertrophy: Long-term results after randomized turbinoplasty}.
\newblock \emph{\bibinfo{journal}{Eur Arch Otorhinolaryngol}}
  \textbf{\bibinfo{volume}{254}}, \bibinfo{pages}{S89--S92}
  (\bibinfo{year}{1997}).

\bibitem{dinis-haider-2002}
\bibinfo{author}{Dinis, P.~B.} \& \bibinfo{author}{Haider, H.}
\newblock \bibinfo{title}{Septoplasty: {{Long-term}} evaluation of results}.
\newblock \emph{\bibinfo{journal}{American Journal of Otolaryngology}}
  \textbf{\bibinfo{volume}{23}}, \bibinfo{pages}{85--90}
  (\bibinfo{year}{2002}).

\bibitem{demirbas-etal-2011}
\bibinfo{author}{Demirbas, D.}, \bibinfo{author}{Cingi, C.},
  \bibinfo{author}{Chakli, H.} \& \bibinfo{author}{Kaya, E.}
\newblock \bibinfo{title}{Use of rhinomanometry in common rhinologic
  disorders}.
\newblock \emph{\bibinfo{journal}{Expert Rev. Med. Devices}}
  \textbf{\bibinfo{volume}{8}}, \bibinfo{pages}{769--777}
  (\bibinfo{year}{2011}).

\bibitem{quadrio-etal-2014}
\bibinfo{author}{Quadrio, M.} \emph{et~al.}
\newblock \bibinfo{title}{Review of computational fluid dynamics in the
  assessment of nasal air flow and analysis of its limitations}.
\newblock \emph{\bibinfo{journal}{Eur Arch Otorhinolaryngol}}
  \textbf{\bibinfo{volume}{271}}, \bibinfo{pages}{2349--2354}
  (\bibinfo{year}{2014}).

\bibitem{radulesco-etal-2019}
\bibinfo{author}{Radulesco, T.} \emph{et~al.}
\newblock \bibinfo{title}{Functional relevance of computational fluid dynamics
  in the field of nasal obstruction: {{A}} literature review}.
\newblock \emph{\bibinfo{journal}{Clin. Otolaryngol.}}
  \textbf{\bibinfo{volume}{44}}, \bibinfo{pages}{801--809}
  (\bibinfo{year}{2019}).

\bibitem{waldmann-etal-2021}
\bibinfo{author}{Waldmann, M.} \emph{et~al.}
\newblock \bibinfo{title}{An effective simulation- and measurement-based
  workflow for enhanced diagnostics in rhinology}.
\newblock \emph{\bibinfo{journal}{Medical \& Biological Engineering \&
  Computing}}  (\bibinfo{year}{2021}).

\bibitem{radulesco-etal-2020}
\bibinfo{author}{Radulesco, T.} \emph{et~al.}
\newblock \bibinfo{title}{Computational fluid dynamics and septal
  deviations---{{Virtual}} surgery can predict post-surgery results: {{A}}
  preliminary study including two patients}.
\newblock \emph{\bibinfo{journal}{Clinical Otolaryngology}}
  \textbf{\bibinfo{volume}{45}}, \bibinfo{pages}{286--291}
  (\bibinfo{year}{2020}).

\bibitem{calmet-etal-2021}
\bibinfo{author}{Calmet, H.} \emph{et~al.}
\newblock \bibinfo{title}{Computational modelling of nasal respiratory flow}.
\newblock \emph{\bibinfo{journal}{Computer Methods in Biomechanics and
  Biomedical Engineering}} \textbf{\bibinfo{volume}{24}},
  \bibinfo{pages}{440--458} (\bibinfo{year}{2021}).

\bibitem{cherobin-etal-2020}
\bibinfo{author}{Cherobin, G.} \emph{et~al.}
\newblock \bibinfo{title}{Rhinomanometry {{Versus Computational Fluid
  Dynamics}}: {{Correlated}}, but {{Different Techniques}}}.
\newblock \emph{\bibinfo{journal}{Am J Rhinol Allergy}}
  \textbf{\bibinfo{volume}{35}}, \bibinfo{pages}{245--255}
  (\bibinfo{year}{2020}).

\bibitem{kimbell-etal-2012}
\bibinfo{author}{Kimbell, J.} \emph{et~al.}
\newblock \bibinfo{title}{Computed nasal resistance compared with
  patient-reported symptoms in surgically treated nasal airway passages: {{A}}
  preliminary report}.
\newblock \emph{\bibinfo{journal}{Am. J. Rhinology {\textbackslash}\& Allergy}}
  \textbf{\bibinfo{volume}{26}}, \bibinfo{pages}{e94--e98}
  (\bibinfo{year}{2012}).

\bibitem{osman-etal-2016}
\bibinfo{author}{Osman, J.} \emph{et~al.}
\newblock \bibinfo{title}{Assessment of nasal resistance using computational
  fluid dynamics}.
\newblock \emph{\bibinfo{journal}{Current Directions in Biomedical
  Engineering}} \textbf{\bibinfo{volume}{2}}, \bibinfo{pages}{617--621}
  (\bibinfo{year}{2016}).

\bibitem{berger-etal-2021}
\bibinfo{author}{Berger, M.} \emph{et~al.}
\newblock \bibinfo{title}{Agreement between rhinomanometry and computed
  tomography-based computational fluid dynamics}.
\newblock \emph{\bibinfo{journal}{Int J CARS}} \textbf{\bibinfo{volume}{16}},
  \bibinfo{pages}{629--638} (\bibinfo{year}{2021}).

\bibitem{schmidt-etal-2022}
\bibinfo{author}{Schmidt, N.}, \bibinfo{author}{Behrbohm, H.},
  \bibinfo{author}{Goubergrits, L.}, \bibinfo{author}{Hildebrandt, T.} \&
  \bibinfo{author}{Br{\"u}ning, J.}
\newblock \bibinfo{title}{Comparison of rhinomanometric and computational fluid
  dynamic assessment of nasal resistance with respect to measurement accuracy}.
\newblock \emph{\bibinfo{journal}{Int J CARS}}  (\bibinfo{year}{2022}).

\bibitem{karbowski-etal-2023}
\bibinfo{author}{Karbowski, K.}, \bibinfo{author}{Kopiczak, B.},
  \bibinfo{author}{Chrzan, R.}, \bibinfo{author}{Gawlik, J.} \&
  \bibinfo{author}{Szaleniec, J.}
\newblock \bibinfo{title}{Accuracy of virtual rhinomanometry}.
\newblock \emph{\bibinfo{journal}{Polish Journal of Medical Physics and
  Engineering}} \textbf{\bibinfo{volume}{29}}, \bibinfo{pages}{59--72}
  (\bibinfo{year}{2023}).

\bibitem{schillaci-quadrio-2022}
\bibinfo{author}{Schillaci, A.} \& \bibinfo{author}{Quadrio, M.}
\newblock \bibinfo{title}{Importance of the numerical schemes in the {{CFD}} of
  the human nose}.
\newblock \emph{\bibinfo{journal}{Journal of Biomechanics}}
  \textbf{\bibinfo{volume}{138}}, \bibinfo{pages}{111100}
  (\bibinfo{year}{2022}).

\bibitem{johnsen-2024}
\bibinfo{author}{Johnsen, S.}
\newblock \bibinfo{title}{Computational {{Rhinology}}: {{Unraveling
  Discrepancies}} between {{In Silico}} and {{In Vivo Nasal Airflow
  Assessments}} for {{Enhanced Clinical Decision Support}}}.
\newblock \emph{\bibinfo{journal}{Bioengineering}}
  \textbf{\bibinfo{volume}{11}}, \bibinfo{pages}{239} (\bibinfo{year}{2024}).

\bibitem{vogt-etal-2016}
\bibinfo{author}{Vogt, K.}, \bibinfo{author}{Wernecke, K.-D.},
  \bibinfo{author}{Behrbohm, H.}, \bibinfo{author}{Gubisch, W.} \&
  \bibinfo{author}{Argale, M.}
\newblock \bibinfo{title}{Four-phase rhinomanometry: A multicentric
  retrospective analysis of 36,563 clinical measurements}.
\newblock \emph{\bibinfo{journal}{European Archives of Oto-Rhino-Laryngology}}
  \textbf{\bibinfo{volume}{273}}, \bibinfo{pages}{1185--1198}
  (\bibinfo{year}{2016}).

\bibitem{hoerschler-schroeder-meinke-2010}
\bibinfo{author}{H{\"o}rschler, I.}, \bibinfo{author}{Schr{\"o}der, W.} \&
  \bibinfo{author}{Meinke, M.}
\newblock \bibinfo{title}{On the assumption of steadiness of nasal cavity
  flow}.
\newblock \emph{\bibinfo{journal}{Biomechanics}} \textbf{\bibinfo{volume}{43}},
  \bibinfo{pages}{1081--1085} (\bibinfo{year}{2010}).

\bibitem{inthavong-etal-2019}
\bibinfo{author}{Inthavong, K.}, \bibinfo{author}{Das, P.},
  \bibinfo{author}{Singh, N.} \& \bibinfo{author}{Sznitman, J.}
\newblock \bibinfo{title}{In silico approaches to respiratory nasal flows:
  {{A}} review}.
\newblock \emph{\bibinfo{journal}{J. Biomech}} \textbf{\bibinfo{volume}{97}},
  \bibinfo{pages}{109434} (\bibinfo{year}{2019}).

\bibitem{quadrio-etal-2016}
\bibinfo{author}{Quadrio, M.} \emph{et~al.}
\newblock \bibinfo{title}{Effect of {{CT}} resolution and radiodensity
  threshold on the {{CFD}} evaluation of nasal airflow}.
\newblock \emph{\bibinfo{journal}{Med Biol Eng Comput}}
  \textbf{\bibinfo{volume}{54}}, \bibinfo{pages}{411--419}
  (\bibinfo{year}{2016}).

\bibitem{tretiakow-etal-2020}
\bibinfo{author}{Tretiakow, D.}, \bibinfo{author}{Tesch, K.},
  \bibinfo{author}{{Meyer-Szary}, J.}, \bibinfo{author}{Markiet, K.} \&
  \bibinfo{author}{Skorek, A.}
\newblock \bibinfo{title}{Three-dimensional modeling and automatic analysis of
  the human nasal cavity and paranasal sinuses using the computational fluid
  dynamics method}.
\newblock \emph{\bibinfo{journal}{Eur Arch Otorhinolaryngol}}
  \textbf{\bibinfo{volume}{278}}, \bibinfo{pages}{1443--1453}
  (\bibinfo{year}{2020}).

\bibitem{tauwald-etal-2024}
\bibinfo{author}{Tauwald, S.} \emph{et~al.}
\newblock \bibinfo{title}{Tomo-{{PIV}} in a patient-specific model of human
  nasal cavities: A methodological approach}.
\newblock \emph{\bibinfo{journal}{Meas. Sci. Technol.}}
  \textbf{\bibinfo{volume}{35}}, \bibinfo{pages}{055203}
  (\bibinfo{year}{2024}).

\bibitem{menter-kuntz-langtry-2003}
\bibinfo{author}{Menter, F.}, \bibinfo{author}{Kuntz, M.} \&
  \bibinfo{author}{Langtry, R.}
\newblock \bibinfo{title}{Ten {{Years}} of {{Industrial Experience}} with the
  {{SST Turbulence Model}}}.
\newblock \emph{\bibinfo{journal}{Turbulence, Heat and Mass Tansfer}}
  \textbf{\bibinfo{volume}{4}}, \bibinfo{pages}{8} (\bibinfo{year}{2003}).

\bibitem{luchini-2016}
\bibinfo{author}{Luchini, P.}
\newblock \bibinfo{title}{Immersed-boundary simulations of turbulent flow past
  a sinusoidally undulated river bottom}.
\newblock \emph{\bibinfo{journal}{Eur. J. Mech. B/Fluids}}
  \textbf{\bibinfo{volume}{55}}, \bibinfo{pages}{340--347}
  (\bibinfo{year}{2016}).

\bibitem{luchini-etal-2025}
\bibinfo{author}{Luchini, P.} \emph{et~al.}
\newblock \bibinfo{title}{A simple and efficient immersed-boundary method for
  the incompressible {{Navier--Stokes}} equations}.
\newblock \emph{\bibinfo{journal}{J. Comp. Phys.}}
  \textbf{\bibinfo{volume}{(Submitted)}} (\bibinfo{year}{2025}).

\bibitem{russo-luchini-2017}
\bibinfo{author}{Russo, S.} \& \bibinfo{author}{Luchini, P.}
\newblock \bibinfo{title}{A fast algorithm for the estimation of statistical
  error in {{DNS}} (or experimental) time averages}.
\newblock \emph{\bibinfo{journal}{J. Comput. Phys.}}
  \textbf{\bibinfo{volume}{347}}, \bibinfo{pages}{328--340}
  (\bibinfo{year}{2017}).

\end{thebibliography}
\end{document}